\documentclass[showpacs,pra,twocolumn,aps,amsmath,amssymb]{revtex4}
\usepackage{color}
\usepackage{graphicx}
\usepackage{dcolumn}
\usepackage{bm}
\newcommand{\bra}[1]{\langle#1|}
\newcommand{\ket}[1]{|#1\rangle}

\newcommand{\scal}[2]{\langle#1|#2\rangle}
\providecommand{\openone}{\leavevmode\hbox{\small1\kern-3.8pt\normalsize1}}

\begin{document}

\title{Distributed correlations and information flows within a hybrid multipartite quantum-classical system}

\author{Bruno Leggio,$^{1}$ Rosario Lo Franco,$^{2,3,4}$ Diogo O. Soares-Pinto,$^{3}$ Pawe{\l} Horodecki,$^5$ and Giuseppe Compagno$^2$}
\affiliation{$^1$Laboratoire Charles Coulomb (L2C), UMR 5221 CNRS-Universit\'{e} de Montpellier - F- 34095 Montpellier, France\\
$^2$Dipartimento di Fisica e Chimica, Universit\`{a} di Palermo, via Archirafi 36, 90123 Palermo, Italy\\
$^3$Instituto de F{\'{i}}sica de S{\~{a}}o Carlos, Universidade de S{\~{a}}o Paulo, CP 369, 13560-970 S{\~{a}}o Carlos, SP, Brasil\\
$^4$School of Mathematical Sciences, The University of Nottingham, University Park, Nottingham NG7 2RD, United Kingdom\\
$^5$Faculty of Applied Physics and Mathematics, Technical University of Gda\`{n}sk, 80-952 Gda\`{n}sk, Poland}

\begin{abstract}
Understanding the non-Markovian mechanisms underlying the revivals of quantum entanglement in the presence of classical environments is central in the theory of quantum information. Tentative interpretations have been given by either the role of the environment as a control device or the concept of hidden entanglement. We address this issue from an information-theoretic point of view. To this aim, we consider a paradigmatic tripartite system, already realized in the laboratory, made of two independent qubits and a random classical field locally interacting with one qubit alone. We study the dynamical relationship between the two-qubit entanglement and the genuine tripartite correlations of the overall system, finding that collapse and revivals of entanglement correspond, respectively, to raise and fall of the overall tripartite correlations. Interestingly, entanglement dark periods can enable plateaux of nonzero tripartite correlations. We then explain this behavior in terms of information flows among the different parties of the system. Besides showcasing the phenomenon of the freezing of overall correlations, our results provide new insights on the origin of retrieval of entanglement within a hybrid quantum-classical system.      
\end{abstract}


\maketitle

\section{Introduction}
Multipartite open quantum systems and the correlations established among their constituent parts play a crucial role in the quantum theory for both fundamental problems and for realistic quantum information processes \cite{benenti,alickibook,horodecki2009RMP,petru,rivasreview,yu2009Science,Modi2012RMP,lofrancoreview}.  
The utilization of quantum correlations, like entanglement, nonlocality and discord, is in fact jeopardized by the detrimental effects of the environment surrounding the quantum system \cite{yu2009Science,Modi2012RMP,lofrancoreview,aolitareview,xu2010NatComm,bellomo2011IJQI,werlangPRA} so that finding strategies to preserve them has become a main requirement. Non-Markovian noise, originating from structured environments or from strong couplings \cite{petru,rivasreview,yu2009Science,fonseca2012}, has been shown to enable dynamical revivals of quantum correlations independently of the quantum \cite{lofrancoreview,bellomo2007PRL,xu2010PRL,bellomoSavasta2011PhyScr,fanchiniPRA} or classical \cite{zhou2010QIP,lofranco2012PhysScripta,bordone2012,lofranco2012PRA,LoFrancoNatCom,altintas2012PLA,benedetti2013,wilsonPRB,bellomo2012noisylaser,
trapani2015,darrigo2013hidden,darrigo2014IJQI,darrigo2012AOP,adeline2014} nature of the environment, allowing an extension of their exploitation time. It is therefore of basic interest to understand the origin of these revival phenomenons.

Non-Markovian dynamics, being linked to memory effects of the environment on the coherence of the quantum system, reveals as a necessary requirement for revivals to occur but it does not provide their interpretation. In the case of local quantum environments, revivals of two-qubit entanglement have been explained by means of periodic entanglement transfers between the qubits and their non-Markovian environments because of the back-action of the environments on the qubits themselves \cite{bellomo2007PRL,Liu2011NatPhys,Lopez2010PRA,chiuri2012,fanchiniSciRep}. Differently, in the case of classical environments, which do not back react on the quantum system and cannot store any quantum correlations, the origin of entanglement revivals appears to be more subtle. Despite this, two all-optical experiments have already confirmed that quantum entanglement can either spontaneously revive \cite{LoFrancoNatCom} or be recovered by local operations \cite{adeline2014} in non-Markovian classical environments.   
Few tentative explanations have been provided so far: one relies on the role played by the classical environment as a control mechanism which, thanks to memory effect, keeps a record for what operation has been applied to the quantum system \cite{lofranco2012PRA,LoFrancoNatCom}; 
another one is based on the concept of hidden entanglement, that is the amount of quantum correlations not revealed by the density matrix description of the system state that can surface by means of local operations \cite{darrigo2012AOP,adeline2014,lofrancoPRB}. 

However, a natural strategy to understand the mechanisms underlying these entanglement revivals in absence of backaction is to perform an in-depth dynamical analysis of the correlation distribution among the parts of such a hybrid quantum-classical composite system, analogously to what is done for quantum dissipative environments \cite{Lopez2010PRA,fanchiniSciRep}. We address this investigation by following an information-theoretic point of view. In particular, we consider a paradigmatic tripartite system, already simulated in the laboratory \cite{LoFrancoNatCom}, made of two independent qubits and a random classical field locally interacting with one qubit alone. We look for a dynamical relationship between the two-qubit entanglement and the genuine tripartite correlations of the overall system. We finally provide physical grounds of this relationship in terms of information flows among the different parties of the system. 

The paper is structured as follows. In Section~\ref{sec:system} we introduce the system under consideration, describing its main features. The study of the dynamics of entanglement and of genuine tripartite correlations of the overall system is done in Section~\ref{sec:tripartite}. In Section~\ref{sec:monogamy} we then report the dynamical study of the information flows within the system. We finally summarize our results and discuss their implications in Section~\ref{sec:conclusion}.

\section{The system}\label{sec:system}
\begin{figure}[tbph]
\begin{center}
\includegraphics[width=0.46\textwidth]{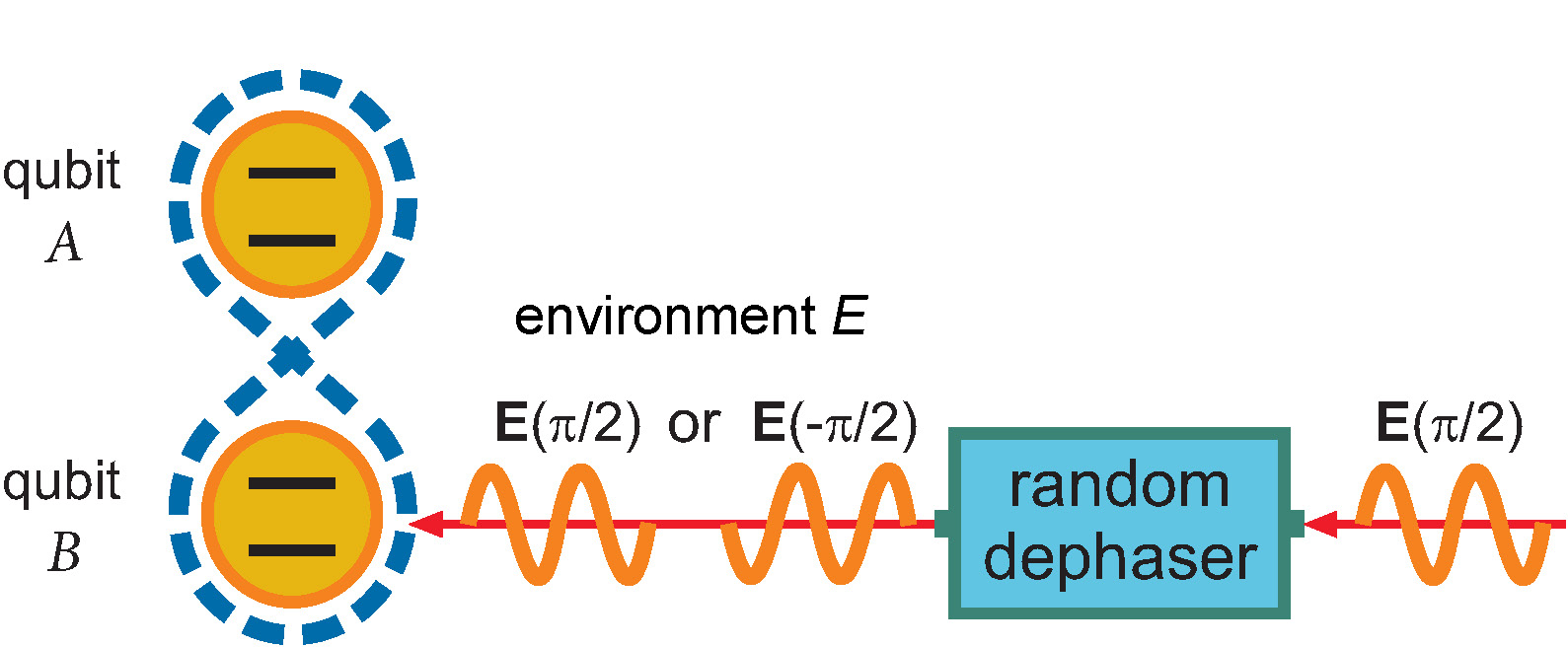}
\end{center}
\caption{\textbf{Scheme of the overall system.} A random external classical field acts on the qubit $B$, whereas qubit $A$ is isolated. The two qubits are initially quantum correlated. The action of the random dephaser is either to shift of $\pi$ with probability $1/2$ the phase ($\pi/2$) of the input field or to leave it unchanged with probability $1/2$.}
\label{fig:system}
\end{figure}

Our system consists of two qubits, one of which (qubit $A$) is isolated while the second one (qubit $B$) interacts with an environment $E$ and, as such, evolves under the action of a non-unitary dynamical map. The two qubits are initially prepared in a quantum correlated state $\rho_{AB}(0)$. As depicted in Fig.~\ref{fig:system}, the environment is a classical field whose phase $\varphi$ is random, being equal to $\varphi_\pm=\pm \frac{\pi}{2}$ with probability $p_{\pm}=\frac{1}{2}$. 

In what follows we use the general notation $\{|0\rangle,|1\rangle\}$ for the computational basis of each qubit. 
The resonant interaction between the qubit $B$ and a classical field $\mathbf{E}$ with phase $\varphi$, in the rotating frame at the qubit-field frequency and within the rotating wave approximation, is represented by the Hamiltonian 
\begin{equation}\label{hamiltonian}
H(\varphi)=\mathrm{i}\hbar (\Omega/2)(\sigma_+e^{-\mathrm{i}\varphi}-\sigma_-e^{\mathrm{i}\varphi}),
\end{equation}
where the qubit-field coupling constant (Rabi frequency) $\Omega$ is proportional to the field amplitude and $\sigma_+=\ket{1}\bra{0}$, $\sigma_-=\ket{0}\bra{1}$ are the raising, lowering operators of the qubit. The time evolution operator $U(t)=\mathrm{e}^{-\mathrm{i}H(\varphi) t/\hbar}$ has the matrix form \cite{lofranco2012PRA,LoFrancoNatCom}
\begin{equation}\label{unitarymatrix}
U_{\varphi,\Omega} (t)=\left(
\begin{array}{cc}\cos(\Omega t/2)&\mathrm{e}^{-\mathrm{i}\varphi}\sin(\Omega t/2)\\
-\mathrm{e}^{\mathrm{i}\varphi}\sin(\Omega t/2) & \cos(\Omega t/2) \\\end{array}\right).
\end{equation}
Considering the random phase of the field and being the qubit $A$ isolated, the state of the bipartite system $A$-$B$ evolves in time according to the map
\begin{equation}\label{rhotnogauss}
\rho_{AB}^\Omega(t)=\frac{1}{2}\sum_{\varphi=\varphi_\pm}
\Big(\openone_A\otimes U_{\varphi,\Omega} (t)\Big)\rho_{AB}(0)\Big(\openone_A\otimes U_{\varphi,\Omega} ^{\dag}(t)\Big),
\end{equation}
where $\openone_A$ is the identity matrix in the Hilbert space of the qubit $A$. We recall that such a map is a complete positive trace preserving one of the class of random unitaries \cite{chruschinskirandom,alickibook}.
A nice feature of this dynamical map is that, if the initial state of the the open system $AB$ at $t=0$ belongs to the class of Bell-diagonal states, the evolved state will remain within such class for each $t>0$. In realistic situations \cite{LoFrancoNatCom}, the field may suffer a noise source due to signal inhomogeneous broadening whose effect is to induce a Gaussian distribution in the field amplitude and thus in the Rabi oscillation frequency $\Omega$ of the qubit evolution, that is 
\begin{equation}\label{gaussian}
G(\Omega_g)=\frac{1}{\sigma \sqrt{\pi}}e^{-\frac{(\Omega_g-\Omega)^2}{4\sigma^2}},
\end{equation}
where $\Omega$ is the Rabi frequency in the absence of dissipation (the central Rabi frequency) and $\sigma$ the Rabi frequency width (standard deviation). 
The evolved state $\rho_{AB}(t)$ in the presence of this field Gaussian noise is obtained by tracing out the Rabi frequency degrees of freedom from 
$\rho_{AB}^\Omega(t)$ of Eq.~\eqref{rhotnogauss}, that is
\begin{equation}\label{rhotgauss}
\rho_{AB}(t)=\int_{-\infty}^{\infty} d\Omega_g \, G(\Omega_g)\, \rho_{AB}^{\Omega_g}(t),
\end{equation} 
with $\int_{-\infty}^{\infty} d\Omega_g \, G(\Omega_g) =1$.
The effect of noise on the random field is therefore transferred to the intrinsic evolution of the quantum system. In the case when the field amplitude is fixed (no Gaussian distribution), the system dynamics is cyclic with no decoherence \cite{lofranco2012PRA,LoFrancoNatCom}.

An advantage of this model is that it can be globally described as a tripartite quantum-classical state, the quantum part played by the two qubits and the classical one by the environment. In particular, including the environmental degrees of freedom of both field phase and field amplitude (that is, Rabi frequency), the classical environment can be represented by means of a state of the form
\begin{equation}\label{environmentstate}
\rho_E=\frac{1}{2}\sum_{\varphi=\varphi_\pm}|\varphi\rangle\langle \varphi| 
\otimes \int_{-\infty}^{\infty} d\Omega_g \, G(\Omega_g)\, \ket{\Omega_g}\bra{\Omega_g},
\end{equation}
which is a classical mixture of the two orthonormal states $\{|\varphi_+\rangle,|\varphi_-\rangle\}$, where $|\varphi_+\rangle$ ($|\varphi_-\rangle$) corresponds to the state of the field with phase $\varphi=\frac{\pi}{2}$ ($\varphi=-\frac{\pi}{2}$), and of the continuous Gaussian variable states $\ket{\Omega_g}$, with 
$\scal{\Omega_g'}{\Omega_g}=\delta(\Omega_g' - \Omega_g)$ and $\delta(x)$ being a Dirac delta function. 
Let us now define a unitary evolution $U_{BE}(t)$ on the bipartition ($BE$) as
\begin{equation}\label{unitaryBE}
U_{BE}(t)=\sum_{\varphi=\varphi_\pm}  \int_{-\infty}^{\infty} d\Omega_g \, U_{\varphi,\Omega} (t) \otimes\ket{\Omega_g}\bra{\Omega_g}
\otimes  |\varphi\rangle\langle \varphi|,
\end{equation}
where $U_{\varphi,\Omega} (t)$ is given in Eq.~\eqref{unitarymatrix}. When there is no Gaussian distribution in the Rabi frequency (periodic dynamics), i.e. $G(\Omega_g)=\delta(\Omega_g-\Omega)$, the environment state and the $B$-$E$ unitary above reduce, respectively, to $\rho_E=\frac{1}{2}\sum_{\varphi=\varphi_\pm}|\varphi\rangle\langle \varphi|$ and $U_{BE}(t)=\sum_{\varphi=\varphi_\pm} U_{\varphi,\Omega} (t) \otimes  |\varphi\rangle\langle \varphi|$. 
For a global initial state $\rho_{ABE}(0)=\rho_{AB}(0)\otimes \rho_E$, the overall evolved state of the tripartite system is then given by
\begin{equation}\label{overallstatet}
\rho_{ABE}(t)=\Big(\openone_A\otimes U_{BE}(t)\Big)\rho_{ABE}(0)\Big(\openone_A\otimes U_{BE}^{\dag}(t)\Big).
\end{equation}
We stress that for a fixed Rabi frequency (delta-like distribution with $\sigma\rightarrow 0$) the global system can be considered closed and the overall dynamics is periodic. Differently, for a Gaussian distribution ($\sigma\neq0$) in the Rabi frequency, that is in the field amplitude, this is to be considered as a further noise source whose continuos degrees of freedom have to be traced out of the overall evolved state above: in this case the global system is open and decays with a decoherence time proportional to $\sigma^{-1}$. 

By performing the suitable partial traces in $\rho_{ABE}(t)$, it is straightforward to obtain the evolved reduced density matrices of all the components (single or bipartite) of the global system. For instance, the two-qubit evolved state $\rho_{AB}(t)$ of Eq.~\eqref{rhotgauss} is retrieved by tracing out the environmental (discrete and continuous) degrees of freedom. It is also possible to show that, at each time during the dynamics, the state of the environment is unaffected by the evolution of the system and, indeed, it does not evolve in time \cite{lofranco2012PRA,LoFrancoNatCom}. This signals the absence of back-action from the environment to the system. Despite this, the dynamics of the open system is non-Markovian. As a matter of fact, any known measure of non-Markovianity such as, for example, the ones given in Refs. \cite{breuer2009PRL,rivas2010PRL}, coincide and witness non-Markovian dynamics when applied to this system \cite{mannone2012,LoFrancoNatCom}. 
Moreover, revivals of entanglement after dark periods occur during the dynamics \cite{lofranco2012PRA,LoFrancoNatCom}. The introduction of the unitary evolution $U_{BE}(t)$ of Eq.~\eqref{unitaryBE} allows a possible interpretation of these revivals by the role of the classical environment as a ``controller'' for which unitary operation is acting on the system. The presence of non-Markovianity (memory effects) in the qubit evolution enables the environment to keep a classical record of what unitary operation has been applied to the qubit $B$. This observation then implies that it is the lack of this classical information that makes entanglement disappear at a given time and it is the recovery of this information that makes quantum correlations then revive. The information the environment holds about a quantum system is therefore due to what action $E$ performs on the system itself, as already discussed within an all-optical experiment for the case without Gaussian frequency distribution \cite{LoFrancoNatCom} and within the context of the so-called hidden entanglement \cite{darrigo2012AOP,adeline2014}.  

The considered model of Fig.~\ref{fig:system} can be described by the standard decoherence paradigm of a quantum system (qubit $A$) quantum correlated with a measurement apparatus (qubit $B$) which in turns interact with an environment ($E$) \cite{zurekreview} and looking for the flows of information between the system $A$ and the environment $E$ \cite{walbornPRL,walbornPRA}. Here we are interested in comprehending the mechanisms which give rise to the revivals of two-qubit entanglement by approaching the problem from an information-theoretic point of view. In particular, we shall investigate how the initial two-qubit quantum correlations are distributed among all the parts of the global system as time goes by, searching for a possible relation between the two-qubit entanglement and the genuine total correlations present in the system. The relevant flows of information present in the overall system shall be then studied.

\section{Tripartite correlations and entanglement}\label{sec:tripartite}
In this section, we shall study the dynamics of two-qubit entanglement and the tripartite correlations present in our global hybrid system, focusing in how these correlations are shared among the three constituents of the system, namely the two qubits and the classical environment. To this end, we shall use a recently-introduced measure of genuine tripartite total correlations \cite{BellomoPRL,mazieroPRA}. 

Given any tripartite system $\{a,b,c\}$, genuine tripartite correlations are, following Ref. \cite{tripartiteconditions}, those which cannot be described as bipartite correlations inside any subsystem $\{i,j\}$ of $\{a,b,c\}$.
The measure $\tau(\rho_{abc})$ of genuine tripartite correlations reads \cite{BellomoPRL,mazieroPRA}
\begin{equation}\label{tripartite}
\tau(\rho_{abc})=\min\big\{I(\rho_{ab,c}),I(\rho_{ac,b}),I(\rho_{bc,a})\big\}.
\end{equation}
It satisfies, on the one hand, the general properties required for such a measure \cite{tripartiteconditions} and has, on the other hand, a very simple expression in terms of bipartite mutual informations 
\begin{equation}\label{MutualInformation}
I(\rho_{ij,k})=S(\rho_{ij})+S(\rho_k)-S(\rho_{ijk}),
\end{equation}
across any possible bipartition $ij$-$k$ of a tripartite system $\{a,b,c\}$. In Eq. \eqref{MutualInformation}, $S(\rho)=-\mathrm{Tr}\rho\ln\rho$ is the von Neumann entropy of the quantum state $\rho$.
The measure $\tau$ of Eq.~\eqref{tripartite} takes into account both classical and quantum correlations of our hybrid quantum-classical system. However, two constituents of this system are quantum objects (the qubits) which can share quantum correlations. Since we are interested in finding a dynamical relation between two-qubit entanglement and tripartite correlations, we quantify the entanglement by the concurrence $\nu(\rho_{AB})$ of the two-qubit reduced state, which is known to be monotonically related to the entanglement of formation for systems of two qubits \cite{horodecki2009RMP}.

\begin{figure}[tbph]
\begin{center}
\includegraphics[width=0.46\textwidth]{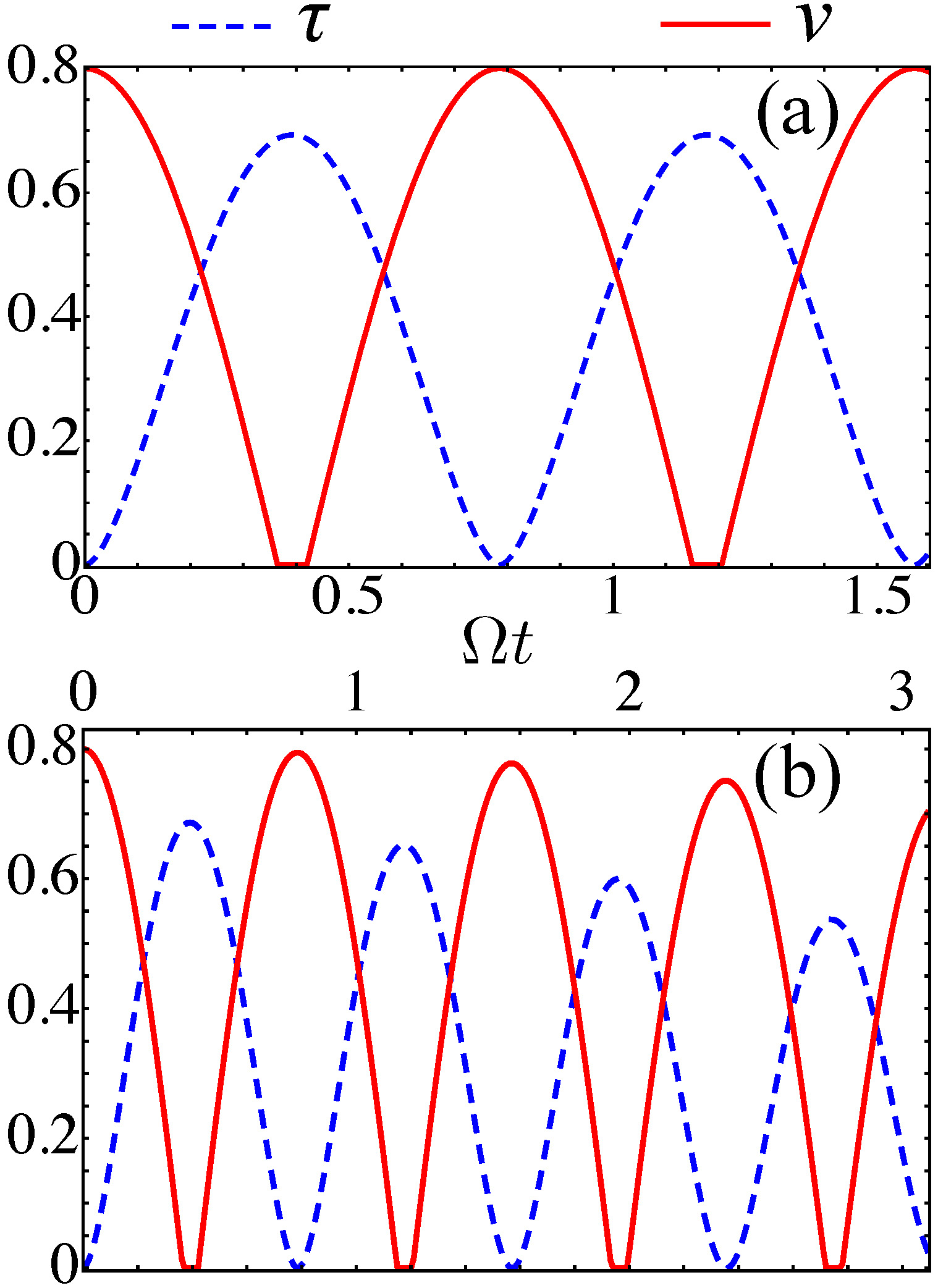}
\end{center}
\caption{Total tripartite correlations $\tau$ in $\rho_{ABE}(t)$ (blue dashed line) and concurrence $\nu$ of $\rho_{AB}(t)$ (red solid line) versus $\Omega t$ for initial conditions $x=z=1$ and $y=0.9$ in the case of \textbf{(a)} periodic dynamics ($\sigma\rightarrow 0$) and \textbf{(b)} decoherent dynamics ($\sigma=0.1\,\Omega$).}
\label{TriEnt1}
\end{figure}

Notice now that, by construction, the two-qubit system and the classical environment are initially decoupled, such that no correlation can be shared between them. The initial amount of correlations present in the overall system therefore depends on the initial state of the two qubits. In order to explore the dynamics originating from different initial conditions, we choose a two-qubit initial state depending on three parameters $x$ ,$y$, $z$ as
\begin{equation}\label{initstate}
\rho_{AB}^0(x,y,z)=y|x_+\rangle\langle x_+|+(1-y)|z_-\rangle\langle z_-|,
\end{equation}
where the state $|\chi_{\pm}\rangle$ ($\chi=x,z$) is defined as
\begin{equation}
|\chi_{\pm}\rangle=\chi |2_{\pm}\rangle+\sqrt{1-\chi^2}|1_{\pm}\rangle,
\end{equation}
$|1_{\pm}\rangle=(\ket{01}\pm\ket{10})/\sqrt{2}$ and $|2_{\pm}\rangle=(\ket{00}\pm\ket{11})/\sqrt{2}$ being, respectively, the Bell (maximally entangled) states in the 1-excitation subspace and in the 0- and 2-excitation subspace. Values of either $x\neq 0,1$ or $z\neq 0,1$ imply an initial linear combination (quantum coherence) between Bell states of different subspaces, while values of $y\neq 0,1$ establish an initial statistical mixture of those states.

We first study the case when the two qubits are initialized in a mixture of Bell states, a Bell-diagonal state, as $\rho_1=\rho_{AB}^0(1,0.9,1)$ having a concurrence $\nu_1=0.8$. Notice that this is the initial state considered in the experiment performed in Ref.~\cite{LoFrancoNatCom}.
The corresponding dynamics of $\nu(t)=\nu\big(\rho_{AB}(t)\big)$ and of $\tau(t)=\tau\big(\rho_{ABE}(t)\big)$ are displayed in Fig. \ref{TriEnt1} by, respectively, red solid and blue dashed lines. Two cases are reported there: in the first one, Panel \textbf{(a)}, the qubit-field coupling is taken as fixed (that is, zero Gaussian standard deviation) so that the evolution of the state $\rho_{ABE}$ is periodic; in the second one, (Panel \textbf{(b)}), the Gaussian distribution of the Rabi frequency (qubit-field coupling) of Eq.~\eqref{gaussian} is considered. Throughout this paper, we have chosen $\sigma=0.1\,\Omega$. The periodic (non-decoherent) dynamics can be viewed as the dynamics of the system when observed at times much shorter than the (Gaussian-induced) decoherence time. 
As expected, the two qubits are initially entangled and uncorrelated with the field. The qubit-field coupling, then, reduces entanglement in time while correlating the environment with the two-level systems. However, correlations in the overall system do not flow from the subsystem $\{A,B\}$ (as entanglement) to the subsystem $\{B,E\}$ (as classical correlations), as one might expect given the fact that only qubit $B$ interacts with $E$. 
In fact it is possible to show, by tracing out the qubit $A$ from the global evolved state of Eq.~\eqref{overallstatet}, that $B$ and $E$ always remain uncorrelated during the dynamics.
They rather turn into genuine tripartite correlations, as clearly shown in Fig. \ref{TriEnt1}. As entanglement decreases, genuine tripartite correlations are built in time: $\nu$ and $\tau$ show a striking ``phase-opposition'' time-behavior, such that the maxima of $\tau$ coincide with the minima of $\nu$ and viceversa. Nevertheless, despite the periodic sudden death and birth of entanglement (due to the initial mixedness of $\rho_{AB}$), $\tau$ has a smooth time-behavior.

\begin{figure}[tbph]
\begin{center}
\includegraphics[width=0.46\textwidth]{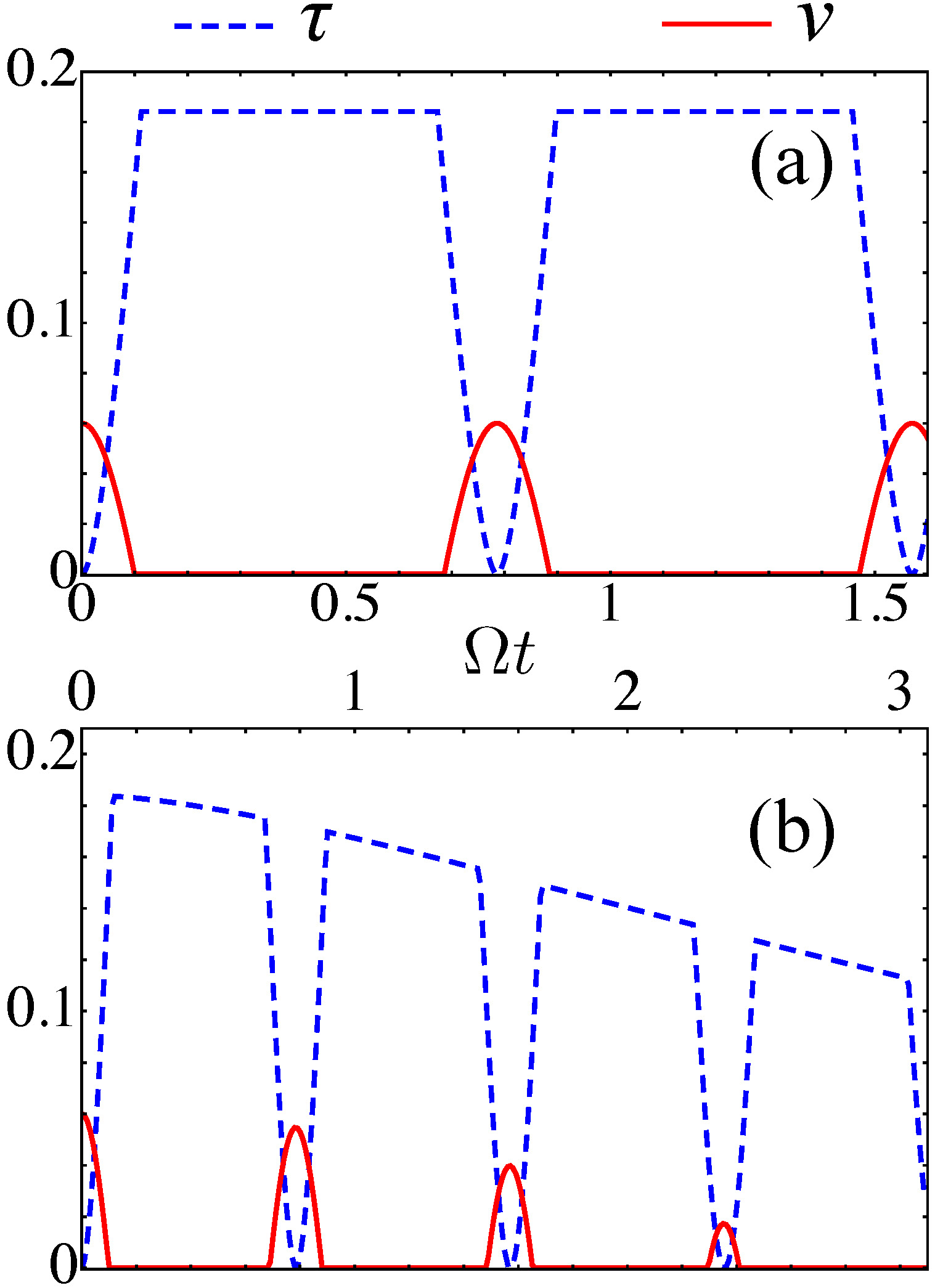}
\end{center}
\caption{Total tripartite correlations $\tau$ in $\rho_{ABE}(t)$ (blue dashed line) and concurrence $\nu$ of $\rho_{AB}(t)$ (red solid line) versus $\Omega t$ for initial conditions $x=0.6, y=0.8$ and $z=0.3$ in the case of \textbf{(a)} periodic dynamics ($\sigma\rightarrow 0$) and \textbf{(b)} decoherent dynamics ($\sigma=0.1\,\Omega$).}
\label{TriEnt2}
\end{figure}

This is no longer the case if initial coherence between Bell states is introduced. In Fig. \ref{TriEnt2} the dynamics of $\nu(t)$ and $\tau(t)$ is shown for the qubits initially in the state $\rho_2=\rho_{AB}^0(0.6,0.8,0.3)$. 
The dynamical phase-opposition between two-qubit entanglement and tripartite correlations is maintained, while a new trait occurs: genuine tripartite correlations freeze for finite time periods, showing a plateau in correspondence of the plateau of zero entanglement (Fig.~\ref{TriEnt2}(a)). 

The understanding of this freezing and whether it also implies a freezing of other kinds of shared correlations within the overall system, will be addressed in the next Section.

\section{Monogamy of correlations: a physical picture of the freezing}\label{sec:monogamy}
Monogamy is a well-known property of various kinds of correlations, either classical or quantum. 
Recently, a monogamy relation for the mutual information in a tripartite system $\{a,b,c\}$, involving genuine tripartite correlations $\tau$, has been given as \cite{costaPRA}
\begin{equation}\label{monogamy}
\tau=\mathcal{I}-\mathcal{I}_{\mathrm{LOC}}-\mu_2,
\end{equation}
where
\begin{equation}
\mu_2=\max_{\mathrm{two-party}}\{I(\rho_{i,j})\},
\end{equation}
is the maximal mutual information obtained over any possible two-party reduced states $\rho_{ij}$,
\begin{equation}\label{stateinfo}
\mathcal{I}=\mathcal{I}(\rho_{abc})=\ln d-S(\rho_{abc}),
\end{equation}
is the so-called state information of the total tripartite state $\rho_{abc}$ living in a Hilbert space of dimension $d$, and
\begin{equation}
\mathcal{I}_{\mathrm{LOC}}=\mathcal{I}(\rho_{a})+\mathcal{I}(\rho_{b})+\mathcal{I}(\rho_{c}),
\end{equation}
is the total state information stored \textit{locally} in each party. Notice that, being Eq.~\eqref{stateinfo} valid for a generic system of state $\rho$, one has $\mathcal{I}(\rho_{i})=\ln d_i-S(\rho_{i})$ ($i=a,b,c$). Eq. \eqref{monogamy} can also be interpreted as follows: the total information stored in a three-party state is the sum of the information stored locally in each party, of the maximal bipartite information in the system and of its genuine tripartite correlations. Local information, bipartite and tripartite correlations thus constitute three boxes where the system can store its total information. 

\begin{figure}[tbph]
\begin{center}
\includegraphics[width=0.46\textwidth]{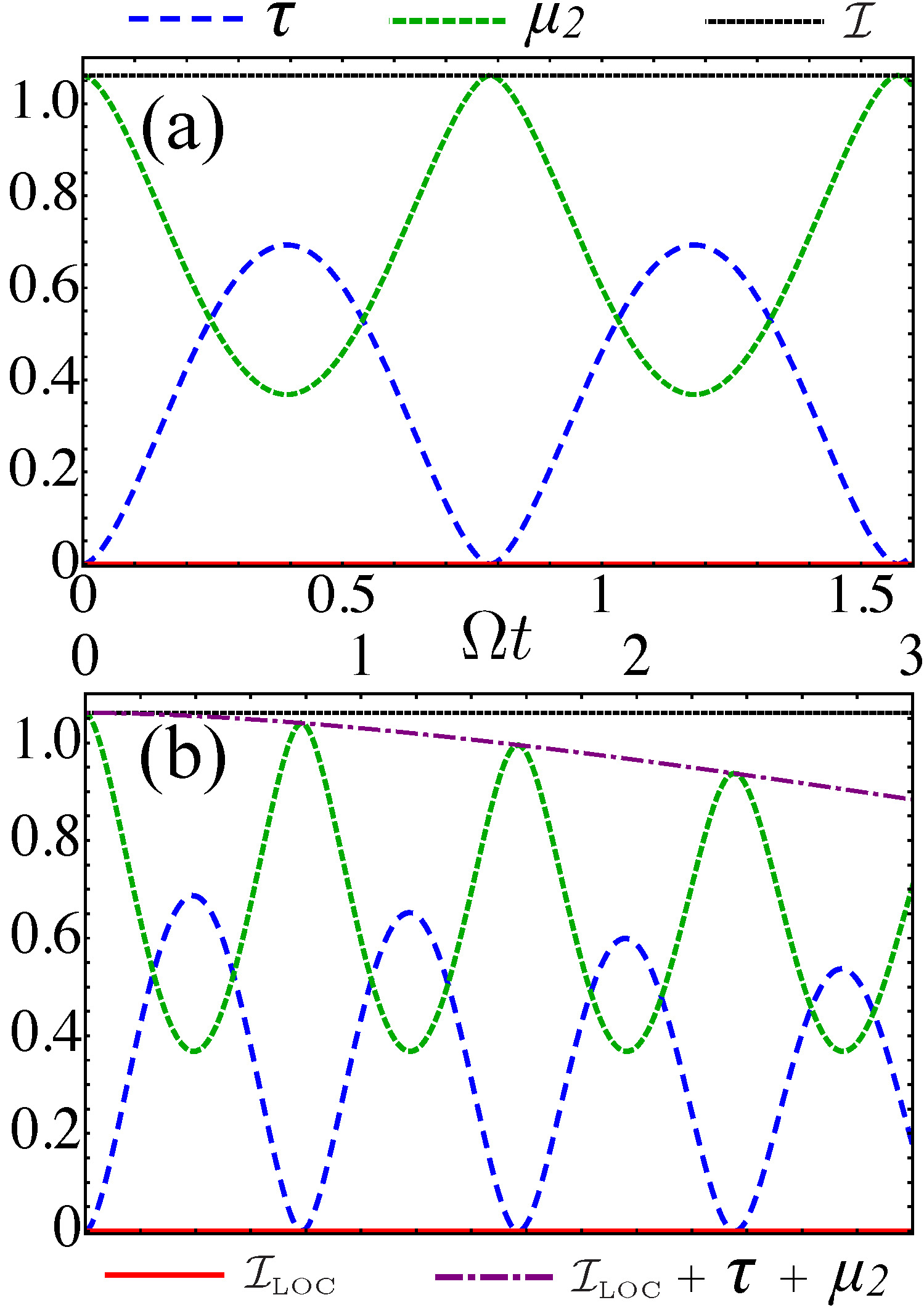}
\end{center}
\caption{Genuine tripartite correlations $\tau$ (dashed blue line), total state information $\mathcal{I}$ (dotted black line), maximal bipartite correlations $\mu_2$ (green dashed line) and local state information $\mathcal{I}_{\mathrm{LOC}}$ (red solid line) versus $\Omega t$ for initial conditions $x=1, y=0.9$ and $z=1$ in the case of \textbf{(a)} periodic dynamics ($\sigma\rightarrow 0$) and \textbf{(b)} decoherent dynamics ($\sigma=0.1\,\Omega$). In this last case, the dot-dashed purple line represents the time-dependent total state information $ \mathcal{I}(t)=\tau+\mathcal{I}_{\mathrm{LOC}}+\mu_2$ while the dotted black line is the initial state information $\mathcal{I}\equiv \mathcal{I}(0)$.}
\label{Monog1}
\end{figure}
In Fig. \ref{Monog1} we show the dynamics of all the quantities involved in Eq. \eqref{monogamy} for our tripartite system, when the two qubits are initially in the state $\rho_1$ defined in the previous Section. We point out that, while in the case of periodic dynamics (closed system) $\mathcal{I}\equiv \mathcal{I}(0)$ is constant, in the decoherent (Gaussian-induced) dynamics the total state information $\mathcal{I}(t)=\tau+\mathcal{I}_{\mathrm{LOC}}+\mu_2$ decays, as shown in panel \textbf{(b)} of Fig.~\ref{Monog1}. 
A particular feature of the dynamics in this case is that the local state information $\mathcal{I}_{\mathrm{LOC}}$ is constantly zero. The information about the total state is always stored in bipartite and-or tripartite correlations. In particular, the information is periodically transferred back and forth between bipartite and tripartite correlations.

\begin{figure}[tbph]
\begin{center}
\includegraphics[width=0.46\textwidth]{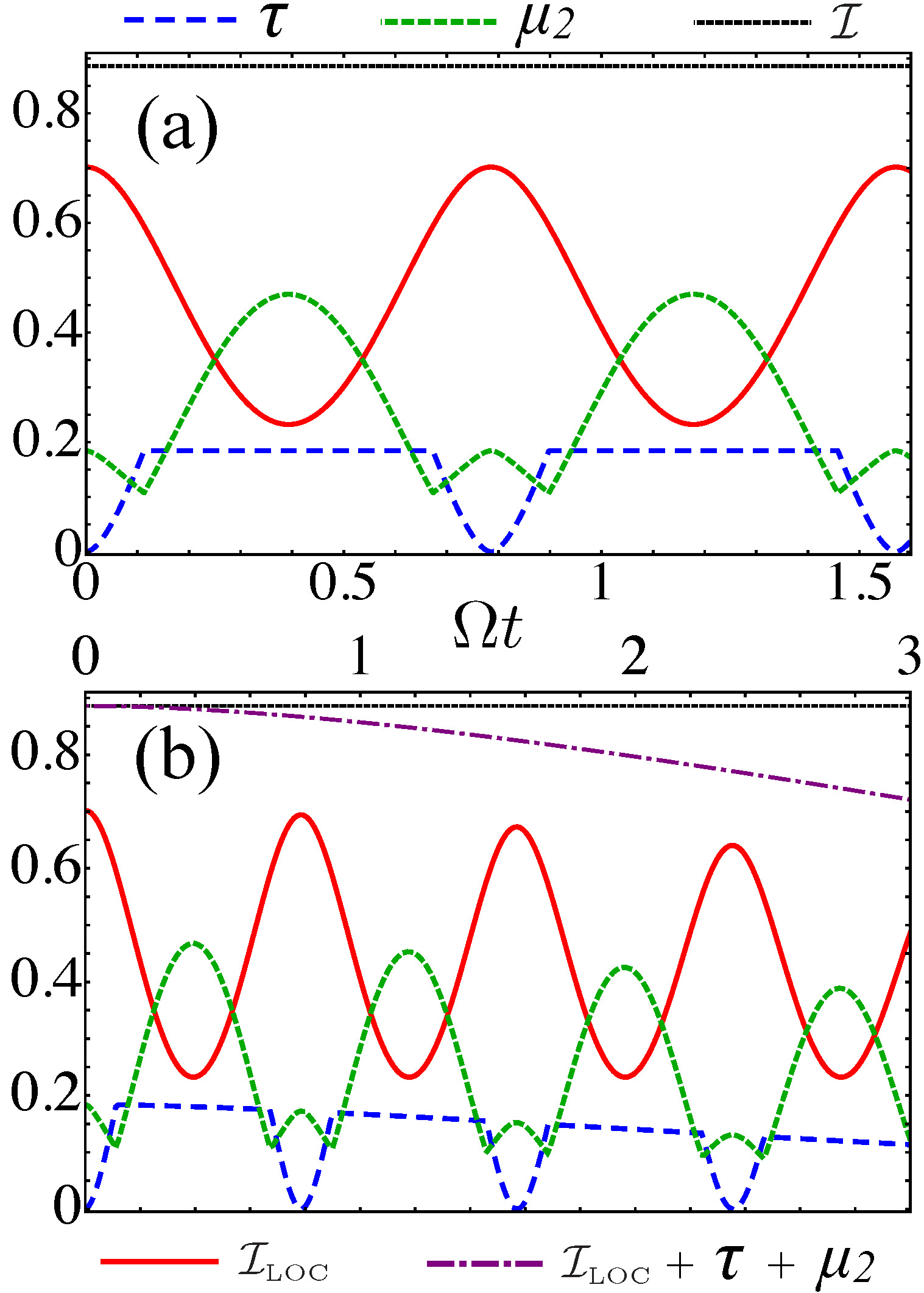}
\end{center}
\caption{Genuine tripartite correlations $\tau$ (dashed blue line), total state information $\mathcal{I}$ (dotted black line), maximal bipartite correlations $\mu_2$ (green dashed line) and local state information $\mathcal{I}_{\mathrm{LOC}}$ (red solid line) versus $\Omega t$ for initial conditions $x=0.6, y=0.8$ and $z=0.3$ in the case of \textbf{(a)} periodic dynamics ($\sigma\rightarrow 0$) and \textbf{(b)} decoherent dynamics ($\sigma=0.1\,\Omega$).  In this last case, the dot-dashed purple line represents the time-dependent total state information $\mathcal{I}(t)=\tau+\mathcal{I}_{\mathrm{LOC}}+\mu_2$ while the dotted black line is the initial state information $\mathcal{I}\equiv \mathcal{I}(0)$.}
\label{Monog2}
\end{figure}
We now examine the case when the qubits are initially in the state $\rho_2$ (see Sec.~\ref{sec:tripartite}). The corresponding dynamics of Eq. \eqref{monogamy} is shown in Fig. \ref{Monog2}. The presence of initial coherence between Bell states produces three more interesting features. First, the local information is now nonzero and oscillates periodically in time. Second, and more importantly, $\mu_2(t)$ can now exceed $\mu_2(0)$ during the evolution. Finally, the freezing of tripartite correlations occurs exactly at times $t_k$ such that $\tau(t_k)=\mu_2(0)$, i.e. when all the initial information stored in bipartite correlations has been transferred to genuine tripartite ones.

In order to comprehend what happens at these points, let us take a step back and re-analyze the setup of our system. Despite being a three-party system, it is constituted by two fundamentally different subsystems: the two qubits, quantum objects, and the classical environment. It represents a hybrid quantum-classical system. We further stress that the reduced state of the classical environment is time invariant, always being the maximally mixed state $\rho_E=\frac{1}{2}\sum_{\varphi=\varphi_\pm}|\varphi\rangle\langle \varphi| $ of Eq.~\eqref{environmentstate} after tracing out the Rabi frequency degrees of freedom and thus giving zero local state information ($\mathcal{I}(\rho_{E})=0)$: the red line in Fig.~\ref{Monog2} in fact describes the local information $\mathcal{I}_{\mathrm{LOC}}$ due to the two qubits.
Genuine tripartite correlations $\tau$ involve by definition all the three parties of the system, so that they represent the information shared among qubit $A$, qubit $B$ and environment $E$. 
On the other hand, the maximal bipartite correlations $\mu_2$ are a hybrid quantity since they may or may not involve the subsystem $E$ due to the maximization over any possible two-party reduced states.
From these considerations, one is led to conclude that $\mathcal{I}_{\mathrm{LOC}}$ and $\tau$ are two different and non-mixable forms of information stored in the system: since $\mathcal{I}_{\mathrm{LOC}}$ is independent of $E$ while $\tau$ always depends on it, they cannot transform into each other and do not thus directly affect their respective time behaviors. Nevertheless, $\mathcal{I}_{\mathrm{LOC}}$ can certainly affect the dynamics of $\mu_2$ since local information on the state of each qubit can evolve in time into bipartite qubit-qubit or qubit-field correlations, which enter the definition of $\mu_2$. At the same time, bipartite correlations involving the environment, such as $I(\rho_{AE})$, can affect the dynamics of $\tau$ by being turned into genuine tripartite ones. To sum up,
\begin{equation}\label{flows}
\mathcal{I}_{\mathrm{LOC}}\rightleftarrows \mu_2\rightleftarrows \tau,
\end{equation}
but
\begin{equation}\label{flows1}
\mathcal{I}_{\mathrm{LOC}} \nrightarrow \tau.
\end{equation}
With this in mind, let us analyze Figs. \ref{Monog1} and \ref{Monog2}. 

In Fig. \ref{Monog1}, where $\mathcal{I}_{\mathrm{LOC}}$ is absent, all the information is stored in correlations which, periodically, change from bipartite to tripartite form according to the second side of Eq. \eqref{flows}.
In the process of Fig. \ref{Monog2}, both sides of Eq. \eqref{flows} are simultaneously active. In particular, at $t=0$ information is stored locally in the states of the two qubits \textit{and also} shared in the form of bipartite correlations between the qubits. As time goes by, two effects arise at the same time: the initial bipartite correlations are turned into genuine tripartite ones, and the initial local information is turned into bipartite correlations: the initial decrease of $\mu_2$ is the result of these two effects. When all the initial bipartite correlations have flown into $\tau$, the only active mechanism is that producing the flux $\mathcal{I}_{\mathrm{LOC}}\rightleftarrows\mu_2$, which leaves $\tau$ unaltered. During this second part of the dynamics (within one period), $\tau$ thus stays constant exhibiting the freezing already witnessed in Section \ref{sec:tripartite}; $\mu_2$ instead increases as there is no ``sink'' where bipartite correlations can flow to and its value can become larger than $\mu_2(0)$, provided enough $\mathcal{I}_{\mathrm{LOC}}$ is present.

To have a better quantitative understanding of this phenomenon, we study the structure of the three quantities $\frac{d \mathcal{I}_{\mathrm{LOC}}}{dt}, \frac{d \tau}{dt}$ and $\frac{d \mu_2}{dt}$. In particular, we write down their explicit expressions for the (non-decoherent) case depicted in Fig. \ref{Monog2}(a) during the two time intervals $T^{(1)}=[0,t^*]$ and $T^{(2)}=[t^*,t_M]$, $t^*$ being the time instant when the freezing of $\tau$ starts and $t_M$ the instant of the first maximum of $\mu_2$. It is here convenient to use the notation $q^{(k)}(t)$ to indicate the evolution of the quantity $q$ in the interval $T^{(k)}$. Thus, for instance, 
$\tau^{(1)}(t)$ is to be understood as the time evolution of the genuine tripartite correlations $\tau$ in the interval $T^{(1)}$.
Seeing that $\frac{d}{dt}S(\rho_A)=\frac{d}{dt}S(\rho_E)=0$ (qubit $A$ is isolated and the state of $E$ is time invariant) and $\frac{d}{dt}S(\rho_{ABE})=\frac{d}{dt}S(\rho_{BE})=0$ (both systems are isolated), one has
\begin{equation}\begin{split}\label{fluxes1}
\frac{d\mathcal{I}_{\mathrm{LOC}}^{(1)}(t)}{dt}&=-\frac{d}{dt}S(\rho_B),\\
\frac{d\mu_2^{(1)}(t)}{dt}&=\frac{d}{dt}S(\rho_B)-\frac{d}{dt}S(\rho_{AB}),\\
\frac{d\tau^{(1)}(t)}{dt}&=\frac{d}{dt}S(\rho_{AB}),
\end{split}\end{equation}
because $\tau^{(1)}=I(\rho_{AB,E})$ and $\mu_2^{(1)}=I(\rho_{AB})$.
Equations \eqref{fluxes1} clearly quantify the information fluxes $\mathcal{I}_{\mathrm{LOC}}\rightarrow \mu_2$ and $\mu_2\rightarrow \tau$, confirming the qualitative processes of Eqs.~\eqref{flows} and \eqref{flows1}. At the beginning of the system dynamics, $\{A,B\}$ is decoupled from $E$. As time goes by, the von Neumann entropy $S(\rho_{AB})$ must increase due to the $B$-$E$ interaction since the two qubits leak information to the environment: bipartite correlations in $\{A,B\}$ turn into genuine tripartite ones.
This process goes on until $t^*$ is reached, when $\tau(t)=\mu_2(0)$. We now notice that, being $\{B,E\}$ noninteracting with $A$, the mutual information $I(\rho_{BE,A})$ is constant in time. Moreover, since at $t=0$ $\rho_{ABE}^0=\rho_{AB}\otimes\rho_E$, one has $I(\rho_{BE,A})(t)=I(\rho_{BE,A}^0)$ and
\begin{equation}\begin{split}
I(\rho_{BE,A}^0)&=S(\rho_{BE}^0)+S(\rho_{A}^0)-S(\rho_{ABE}^0)\\
&=S(\rho_{B}^0)+S(\rho_{E}^0)+S(\rho_{A}^0)-S(\rho_{AB}^0)-S(\rho_{E}^0)\\
&=I(\rho_{AB}^0)=\mu_2(0).
\end{split}\end{equation}
This means that, if the condition $\tau(t)=\mu_2(0)$ is achieved, tripartite correlations do not increase anymore over that value since they cannot exceed the steady mutual information of the bipartition $\{BE,A\}$. Thus, either $\tau(t)$ decreases (similar to the dynamics of Fig. \ref{Monog1}, where the condition $\tau(t)=\mu_2(0)$ is however not reached) or, if other information channels are available as in the dynamics of Fig. \ref{Monog2}, freezes to the value $\mu_2(0)$.

During the second time interval $T^{(2)}$ between $t^*$ and $t_M$, we have $\tau^{(2)}=I(\rho_{BE,A})$ and $\mu_2^{(2)}=I(\rho_{BC})$, so that
\begin{equation}\begin{split}\label{fluxes2}
\frac{d\mathcal{I}_{\mathrm{LOC}}^{(2)}(t)}{dt}&=-\frac{d}{dt}S(\rho_B),\\
\frac{d\mu_2^{(2)}(t)}{dt}&=\frac{d}{dt}S(\rho_B)-\frac{d}{dt}S(\rho_{BE})=\frac{d}{dt}S(\rho_B),\\
\frac{d\tau^{(2)}(t)}{dt}&=0,
\end{split}\end{equation}
and the flow $\mathcal{I}_{\mathrm{LOC}}\rightarrow \mu_2$ continues. After $t_M$, due to the periodicity of the dynamics, all the information fluxes get reversed and the system evolves towards the initial condition by following the same information dynamics backwards.

\section{Conclusion}\label{sec:conclusion}
In this paper we have addressed the problem of understanding the mechanisms underlying entanglement revivals in a composite quantum system interacting with a local classical environment by an information-theoretic approach, that is by analyzing the correlation distribution and information flows within the overall system. To this aim, we have considered a system made of two quantum objects (qubits) and a classical environment (random external field) in an experimentally realistic configuration. 
We have shown such a system to offer a rich scenario thanks to the dynamical interplay of two different kinds of correlations, namely quantum and total ones. In particular, we have found a clear dynamical relationship between two-qubit entanglement and genuine tripartite correlations shared among all the parts of the system. In general, if entanglement is initially present in the two-qubit state, it periodically turns into genuine tripartite correlations, despite the third party being a purely classical object. This general behavior highlights the mechanism of spontaneous entanglement recovery during the system evolution. 

We have moreover shown that, if the initial state of the two qubits exhibits a coherence between Bell states which is a typical quantum feature, a dynamical freezing of tripartite correlations occurs in correspondence to dark periods of entanglement. We remark that, while freezing of purely quantum correlations is known to happen under suitable conditions \cite{mazzolaPRL,xu2010NatComm,auccaisePRL,xu2012review,aaronson2013PRA,aaronson2013NJP,haikkaPRA,universalfreezing}, the freezing phenomenon of total multipartite correlations appears here for the first time. It may be of relevance for quantum information protocols relying on overall correlations in a composite system in the presence of a classical environment.
We have then provided a physical picture of the total correlation freezing, which we argue to stem from the existence of non-trivial time behavior of the information stored locally on the two-qubit state, in turn due to the initial coherence between Bell states. In particular, this freezing is a consequence of monogamy of correlations, inducing a periodic redistribution of the total information contained in the tripartite state into different possible forms, namely, local state information, bipartite correlations and genuine tripartite correlations. We have indeed shown that the system can redistribute in time its bipartite correlations without altering the genuine tripartite ones, just by addressing the local information stored in the state of each of its parts.

The results of this work give new insights on the mechanisms underlying the retrieval of entanglement within a hybrid quantum-classical system. They also motivate further studies on the manipulation of hybrid systems for quantum information purposes \cite{hybridPNAS2015}, for instance by looking for operational interpretations of total tripartite correlations in order to exploit their freezing here found or by design suitable classical fields capable to efficiently preserve quantum correlations.

\begin{acknowledgments}
R.L.F. and D.S.P. acknowledge support by the Brazilian funding agency CAPES [Pesquisador Visitante Especial-Grant No. 108/2012].
\end{acknowledgments}

\end{document}